\documentclass[prd,aps,preprint,superscriptaddress,showkeys,showpacs,
 preprintnumbers,amsmath,amssymb,floatfix]{revtex4-1}
\usepackage{graphicx}
\usepackage{epsfig}
\usepackage{amsfonts}
\usepackage{dsfont}
\usepackage{amsmath}
\usepackage{amssymb}
\usepackage[sort&compress]{natbib}
\usepackage{dcolumn}
\usepackage{multirow}
\usepackage{bigstrut}
\usepackage{mathrsfs}
\usepackage{type1cm}
\usepackage{placeins}
\usepackage{color}
\usepackage{rotating}
\usepackage{slashed}
\usepackage{subfigure}
\usepackage[mathscr]{euscript}
\newcommand{\Ll}{\mathcal{L}}

\newcommand{\Tr}{\mathrm{Tr}}

\begin{document}

\title{Gluonic Lorentz violation and chiral perturbation theory}

\author{J. P. Noordmans}
\affiliation{CENTRA, Departamento de F\'isica, Universidade do Algarve, 8005-139 Faro, Portugal}

\date{\today}
\vspace{3em}

\begin{abstract}
\noindent By applying chiral-perturbation-theory methods to the QCD sector of the Lorentz-violating Standard-Model Extension, we investigate Lorentz violation in the strong interactions. In particular, we consider the CPT-even pure-gluon operator of the minimal Standard-Model Extension. We construct the lowest-order chiral effective Lagrangian for three as well as two light quark flavors. We develop the power-counting rules and construct the heavy-baryon chiral-perturbation-theory Lagrangian, which we use to calculate Lorentz-violating contributions to the nucleon self energy. Using the constructed effective operators, we derive the first stringent limits on many of the components of the relevant Lorentz-violating parameter. We also obtain the Lorentz-violating nucleon-nucleon potential. We suggest that this potential may be used to obtain new limits from atomic-clock or deuteron storage-ring experiments.
\end{abstract}

\maketitle

\section{Introduction}

The detection of an experimental signal corresponding to the breakdown of Lorentz symmetry \cite{Lorentzsym} would be a major discovery and could potentially provide valuable information about a possible theory of quantum gravity. Although no such signal has been detected to date, there is still a large interest in the possibility that Lorentz symmetry might be violated in nature. This is caused by the fact that some proposed models of quantum gravity involve mechanisms that allow for (spontaneous) Lorentz violation (LV) at Planck-scale energies \cite{qgmodels1}. Tiny remnants of such high-energy LV might be detectable at experimentally attainable energies, in particular because there is little experimental background from conventional Lorentz-symmetric (LS) physics for many of the corresponding signals.

Searches for LV are arguably \cite{argumentsforsme} best performed in the context of a realistic effective field theory (EFT) for general LV, called the Standard-Model Extension (SME) \cite{sme}. Its particle-physics Lagrangian contains all possible operators that can be constructed using the conventional standard-model fields, coupled to fixed background tensors. These Lorentz-violating coefficients (LVCs) presumably originate from an underlying fundamental theory that (spontaneously) breaks Lorentz symmetry. Being the most general realistic EFT for Lorentz-symmetry breaking, the SME is also the most general realistic EFT for CPT violation \cite{greenberg}.  

From a phenomenological point of view, the virtue of the SME lies in the fact that it provides a means to explicitly calculate observable signals for Lorentz-symmetry breaking, as well as a way to systematically identify unconstrained regions of the LV parameter space. As a consequence, many stringent constraints on LVCs have been obtained experimentally \cite{datatables}. Particularly successful in this respect are low-energy precision tests of nuclear and hadronic systems, providing severe limits on various effective nucleon and other hadronic parameters for LV. However, since quantum chromodynamics (QCD) is nonperturbative at the relevant energies, deriving direct bounds on the more fundamental quark and gluon parameters that appear in the SME Lagrangian, is complicated. This amounts to a relatively small set of direct bounds on quark and gluon parameters \cite{datatables}.

A promising approach, that is aimed at remedying this situation, is applying the well-established machinery of chiral perturbaton theory ($\chi$PT) \cite{foundation} to the QCD sector of the SME \cite{LVchiPT1, LVchiPT2}. It is similar in spirit to studies of the breaking of parity \cite{Kap93} and of time-reversal symmetry \cite{Mer10}. In this work we extend this approach to the CPT-even pure-gluon sector of the minimal Standard-Model Extension (mSME). The latter is the restriction of the full SME to LV operators with mass dimension $d\leq 4$. In Sec.~\ref{sec:SMElagr}, we will introduce the relevant mSME operator and discuss some pertinent properties of the corresponding LVC: $k_G^{\mu\nu\rho\sigma}$. In Sec.~\ref{sec:chiPTlagr} we construct the induced chiral effective Lagrangian in terms of the degrees of freedom that are relevant below the chiral-breaking scale $\Lambda_\chi \simeq 1\ {\rm GeV}$, i.e. the light mesons and baryons. We will investigate the power-counting rules and introduce the LV heavy-baryon Lagrangian in Sec.~\ref{sec:pc}, which we use to calculate the LV contribution to the nucleon self energy. Using the obtained Lagrangians, in Sec.~\ref{sec:nucleonlimits}, we will obtain the first bounds on eight of the nineteen independent components of the LVC from existing bounds on effective neutron and proton parameters. The power of the chiral perturbation approach is exemplified by the observation that the remaining ten components of $k_G^{\mu\nu\rho\sigma}$ do not induce any kinetic nucleon terms. This leads us to conclude that nucleon bounds cannot directly constrain these ten parameters to the desired level of accuracy. On the other hand, such bounds can be obtained by considering contributions of $k_G^{\mu\nu\rho\sigma}$ to a pure-photon operator. This will be considered in Sec.~\ref{sec:photonlimits}. In Sec.~\ref{sec:improvements} we show that additional and/or improved bounds might be obtained by considering Cherenkov-like pion emission by protons and LV pion exchange between nucleons and its effect on for example the spin precession of the deuteron. Finally, in Sec.~\ref{sec:conclusion}, we will summarize and present our conclusions.

\section{The pure gluon CPT-even mSME Lagrangian}
\label{sec:SMElagr}

In the mSME there is one LV CPT-even coefficient that couples to a pure gluon operator. This operator has mass-dimension four and is given by \cite{sme}
\begin{equation}
\Ll = -\frac{1}{2} k_G^{\mu\nu\rho\sigma}\Tr\left(G_{\mu\nu}G_{\rho\sigma}\right) \ ,
\label{SMElagr}
\end{equation}
where $G_{\mu\nu} = \tfrac{1}{2}G_{\mu\nu}^a \lambda^a$ is the gauge field strength of the $SU(3)$ color gauge group (here, $\lambda^a/2$, with $a=1,\ldots,8$, are the corresponding generators) and $k_G^{\mu\nu\rho\sigma}$ is a real tensor that parametrizes the LV. The operator is even under charge conjugation and after either a parity or a time-reversal transformation, it gains a factor $(-1)^\mu (-1)^\nu (-1)^\rho (-1)^\sigma$, with $(-1)^\mu = 1$ if $\mu=0$ and $(-1)^\mu = -1$ otherwise.

In addition to this CPT-even operator there is one CPT-odd gluon operator of mass-dimension three in the mSME. However, it is associated with negative and imaginary contributions to the energy. Although it has recently been shown that consistent quantization is nevertheless possible for the analogous photon parameter upon introducing an unobservably small photon mass \cite{covquantphot}, we will ignore the CPT-odd gluon term in the present case.

The coefficient $k_G^{\mu\nu\rho\sigma}$ in Eq.~\eqref{SMElagr} is real and has the symmetries of the Riemann curvature tensor, i.e. 
\begin{equation}
k_G^{\mu\nu\rho\sigma} = -k_G^{\mu\nu\sigma\rho} = k_G^{\rho\sigma\mu\nu}\ ,\qquad k_G^{\mu[\nu\rho\sigma]} = 0\ ,
\label{symmetries}
\end{equation}
where the square brackets indicate total anti-symmetrization of the enclosed indices. The second relation holds for anti-symmetrization of any group of three indices.  It follows from the fact that, using the first two relations in Eq.~\eqref{symmetries}, we can write $k_G^{\mu[\nu\rho\sigma]} = \frac{1}{24}\epsilon_{\alpha\beta\gamma\delta}k_G^{\alpha\beta\gamma\delta}\epsilon^{\mu\nu\rho\sigma}$. Therefore, a corresponding nonzero part of $k_G^{\mu\nu\rho\sigma}$ does not violate Lorentz symmetry and can be absorbed in the conventional $\bar{\theta}$ term of QCD. Additionally, we can take $k_G^{\mu\nu\rho\sigma}$ to have a vanishing double trace, i.e. $(k_G)^{\mu\nu}_{\;\;\;\;\mu\nu} = 0$, since such a trace part also does not violate Lorentz invariance and can be absorbed in the conventional LS gauge term. 

These considerations show that $k_G^{\mu\nu\rho\sigma}$ has 19 independent real, physical, and LV components. These can be grouped into two groups of 9 and 10 parameters, respectively, by decomposing $k_G^{\mu\nu\rho\sigma}$ as
\begin{equation}
k_G^{\mu\nu\rho\sigma} = E^{\mu\nu\rho\sigma} + W^{\mu\nu\rho\sigma} \ ,
\label{decomposition}
\end{equation}
where
\begin{subequations}
\begin{eqnarray}
E^{\mu\nu\rho\sigma} &=& \frac{1}{2}\left(\eta^{\mu\rho}k^{\nu\sigma}+\eta^{\nu\sigma}k^{\mu\rho} - \eta^{\nu\rho}k^{\mu\sigma} - \eta^{\mu\sigma}k^{\nu\rho}\right)\ , \\
k^{\mu\nu} &=& \eta_{\alpha\beta}k_G^{\mu\alpha\nu\beta}\ , \label{cmunulike}
\end{eqnarray}
\end{subequations}
and $\eta^{\mu\nu}$ is the Minkowski metric tensor. This decomposition is similar to the Ricci decomposition of the Riemann curvature tensor, with $E^{\mu\nu\rho\sigma}$ the semi-traceless part build in terms of the Ricci curvature $k^{\mu\nu}$ and $W^{\mu\nu\rho\sigma}$ the fully traceless Weyl tensor (the would-be Ricci scalar vanishes because $k_G^{\mu\nu\rho\sigma}$ is doubly traceless). A convenient way of writing $E^{\mu\nu\rho\sigma}$ and $W^{\mu\nu\rho\sigma}$ is
\begin{equation}
E^{\mu\nu\rho\sigma} = \frac{1}{2}\left(k_G^{\mu\nu\rho\sigma} + \breve{k}_G^{\mu\nu\rho\sigma}\right)\ ,\qquad W^{\mu\nu\rho\sigma} = \frac{1}{2}\left(k_G^{\mu\nu\rho\sigma} - \breve{k}_G^{\mu\nu\rho\sigma}\right)\ ,
\label{levicivdif}
\end{equation}
with $\breve{k}_G^{\mu\nu\rho\sigma} = \frac{1}{4}\epsilon^{\mu\nu\alpha\beta}\epsilon^{\rho\sigma\gamma\delta}(k_G)_{\alpha\beta\gamma\delta}$ and $\epsilon^{\mu\nu\rho\sigma}$ the Levi-Civita tensor with $\epsilon^{0123} = +1$. Some additional intuition can be gained by comparing $k_G^{\mu\nu\rho\sigma}$ to its $U(1)$ photon analogue, $k_F^{\mu\nu\rho\sigma}$, which has been studied in much greater detail than $k_G^{\mu\nu\rho\sigma}$. Using Eq.~\eqref{levicivdif} it is easy to see that the 10 independent components of $W^{\mu\nu\rho\sigma}$ are the gluon analogues of the birefringent components of $k_F^{\mu\nu\rho\sigma}$, which cause the vacuum to have an effective refractive index \cite{photon1}. In contrast, the 9 independent components of $E^{\mu\nu\rho\sigma}$ should be compared to the non-birefringent part of $k_F^{\mu\nu\rho\sigma}$ \cite{photon1}. Both $E^{\mu\nu\rho\sigma}$ and $W^{\mu\nu\rho\sigma}$ obey Eq.~\eqref{symmetries}. 

Presently, the only reported bound on $k_G^{\mu\nu\rho\sigma}$ is obtained through quantum mixing of $k^{\mu\nu}$ with the LVC $c^{\mu\nu}$ of the electron \cite{boundonkG}. The resulting bound is given by
\begin{equation}
|\tilde{k}_{tr}| = \frac{2}{3}|k^{00}| < 4 \times 10^{-15}\ .
\end{equation}
This leaves 18 of the 19 independent CPT-even mSME pure-gluon parameters unbounded. However, one expects at least some of them to contribute to effective LV parameters for nucleons and hadrons, for which stringent limits have been obtained \cite{datatables}. Therefore, we will consider the EFT of QCD, chiral perturbation theory, which is formulated in terms of these (effective) degrees of freedom.

\section{The effective chiral Lagrangian}
\label{sec:chiPTlagr}

We construct the low-energy effective chiral Lagrangian corresponding to Eq.~\eqref{SMElagr} in the formalism of Gasser and Leutwyler \cite{Gas84}. For a pedagogical introduction see Ref.~\cite{primer}. As any approach to $\chi$PT, it is based on the observation that the QCD Lagrangian, containing only gluons and the lightest three quarks, is approximately invariant under global $SU(3)_L\times SU(3)_R\times U(1)_V$ transformations of the quark fields (disregarding the axial $U(1)_A$ symmetry, which is broken by quantum anomalies). From the absence of parity doubling in the hadron spectrum one deduces that the axial $SU(3)_A$ part of the $SU(3)_L\times SU(3)_R$ symmetry must be spontaneously broken, leaving $SU(3)_V\times U(1)_V$ as the remaining symmetry group. The pseudo-Goldstone bosons associated with the symmetry breaking are identified with the light $J^P = 0^-$ mesons, which have a small mass compared to the $J^P = 1^-$ vector mesons and the $J^P = \tfrac{1}{2}^+$ baryons. The nonzero masses of the pseudoscalar mesons originate from the fact that in the QCD Lagrangian the $SU(3)_A$ symmetry is explicitly broken by the (small) quark masses. 

The effective Lagrangian is constructed in terms of the low-energy degrees of freedom, i.e. the light-meson fields and the baryons, such that it contains all possible operators that obey the symmetries of the QCD Lagrangian \cite{foundation}. To correctly implement these symmetries, the light-meson fields are collected in the unitary matrix
\begin{equation}
U = \exp(i\phi_a(x) \lambda_a/F_0)\ ,
\label{Umatrix}%
\end{equation}
where $\phi_a$ are the pseudo-Goldstone fields, $\lambda_a$ are the Gell-Mann matrices, and $F_0 \simeq \Lambda_\chi/(4\pi)$ is the pion-decay constant in the limit of vanishing quark masses, i.e. the chiral limit. The matrix $U$ transforms as
\begin{equation}
U \rightarrow R U L^\dagger
\label{Uchiraltrafo}
\end{equation}
under chiral transformations. Here, the global matrices $R$ and $L$ are independent $SU(3)$ matrices. We disregard interactions with external fields in this work, since they appear only in higher-order effects (however, see Sec.~\ref{sec:photonlimits}). Consistent introduction of such interactions would require Eq.~\eqref{Uchiraltrafo} to become a local transformation.

To apply the QCD symmetries to the effective baryon Lagrangian, one defines the unitary square root of the matrix $U$ in Eq.~\eqref{Umatrix} by $u$, i.e. $u^2 = U$. The chiral transformation of $u$ leads to the definition of the unitary matrix $K = K(L,R,U)$ by $u \rightarrow u' = \sqrt{R U L^\dagger}  \equiv R u K^{-1}$, or $K = (R U L^\dagger)^{-\tfrac{1}{2}} R u = u'^{\dagger}R u = u' L u^\dagger$. Subsequently, the $\tfrac{1}{2}^+$ baryon octet, described by eight Dirac spinors $B_a$, with $a = 1,\ldots,8$, is represented by the traceless $3\times 3$ matrix 
\begin{equation}
B = \frac{B_a \lambda_a}{\sqrt{2}}\ ,
\label{baryonoctet}
\end{equation}
that transforms under global $SU(3)_L\times SU(3)_R$ as
\begin{equation}
B \rightarrow K B K^\dagger\ .
\end{equation}
The chiral covariant derivative of $B$ is defined using the chiral connection $\Gamma^\mu = \frac{1}{2}\left[u^\dagger\partial^\mu u + u\partial^\mu u^\dagger\right]$ and is given by
\begin{equation}
D_\mu B = \partial_\mu B + \left[\Gamma^\mu , B\right]\ .
\end{equation}
The final building block of the Lagrangian we need is the chiral vielbein, given by 
\begin{equation}
u^\mu = i\left[u^\dagger \partial^\mu u - u \partial^\mu u^\dagger\right]\ ,
\end{equation}
that transforms as $u^\mu \rightarrow K u^\mu K^\dagger$ under chiral transformations. Both $\Gamma^\mu$ and $u^\mu$ become dependent on external fields, when one includes them.

Based on the chiral-transformation properties of the different building blocks of the Lagrangian one builds all operators that have the symmetry properties of the QCD Lagrangian. These operators can be ordered by powers of the expansion parameter $q/\Lambda_\chi$, where $q \sim m_\pi \ll \Lambda_\chi$ is the typical momentum of the process. We will discuss the power counting in more detail in the next section.

The lowest order light-meson and baryon Lagrangians are then given by \cite{Gas84, Geo84, Krau90}
\begin{subequations}
\begin{eqnarray}
\mathcal{L}_{\phi} &=& \frac{F_0^2}{4}{\rm Tr}\left[\partial_\mu U(\partial^\mu U)^\dagger\right] + \frac{F_0^2 B_0}{2}{\rm Tr}\left[\mathscr{M}U^\dagger + U\mathscr{M}^\dagger\right]\ , \label{LOlightmeson} \\
\mathcal{L}_{\phi B} &=& {\rm Tr}\left[\bar{B}(i\slashed{D} - m_0)B\right] + \frac{D}{2}{\rm Tr}\left[\bar{B}\gamma^\mu\gamma^5\{u_\mu,B\}\right] + \frac{F}{2}{\rm Tr}\left[\bar{B}\gamma^\mu\gamma^5\left[u_\mu,B\right]\right]\ ,
\label{LObaryon}%
\end{eqnarray}
\label{LOLS}%
\end{subequations}
respectively. Here, $\mathscr{M} = {\rm diag}(m_u,m_d,m_s)$ is the quark-mass matrix. The mass term in QCD Lagrangian breaks chiral symmetry, However, it would be invariant under chiral transformations if $\mathscr{M}$ would transform as $\mathscr{M} \rightarrow R \mathscr{M} L^\dagger$. This property is mimicked by Eq.~\eqref{LOlightmeson} and it exemplifies how symmetry-breaking terms are incorporated into the formalism \cite{Geo84}. 

Furthermore, $B_0$, $D$, and $F$ are low-energy constants (LECs) whose size cannot be determined using symmetry arguments. However, an order-of-magnitude estimate can be given, using naive dimensional analysis (NDA) \cite{Man84}. In this case NDA gives $B_0 = \mathcal{O}(\Lambda_\chi)$, which leads for example to $m_\pi^2 = \mathcal{O}((m_u + m_d)\Lambda_\chi)$, which agrees fairly well with the actual pion mass. For this reason an insertion of the quark-mass matrix in the Lagrangian counts as $\mathcal{O}(q^2)$ for the power counting. The coefficients $D$ and $F$ are experimentally determined to be $D=0.80$ and $F=0.50$ (at tree level) \cite{borasoy}. This also agrees well with NDA estimates, which give $D,F = \mathcal{O}(1)$.

In the same way the conventional QCD Lagrangian gives rise to a LS low-energy effective Lagrangian, the LV operators in the QCD sector of the SME can be related to effective operators in a LV chiral Lagrangian \cite{LVchiPT1,LVchiPT2}. Being a pure gluon operator, Eq.~\eqref{SMElagr} is trivially invariant under chiral transformations of the quark fields. The lowest-order relevant effective operators that capture this property, as well as the C, P, and T characteristics of Eq.~\eqref{SMElagr}, are given by
\begin{subequations}
\begin{eqnarray}
\mathcal{L}^{k_G}_\phi &=& \frac{F_0^2 r_1}{4} k_{\mu\nu}{\rm Tr}\left[(\partial^\mu U)^\dagger \partial^\nu U\right]\ , \label{LOLVmeson} \\
\mathcal{L}^{k_G}_{\phi B} &=& i r_2 k_{\mu\nu}{\rm Tr}\left[\bar{B}\gamma^\mu D^\nu B\right] +  \frac{ir_3}{2m_0} \tilde{W}_{\mu\nu\rho\sigma}{\rm Tr}\left[\bar{B}\sigma^{\mu\nu}[ u^\rho, D^\sigma B]\right] \notag \\
&& +  \frac{ir_4}{2m_0} \tilde{W}_{\mu\nu\rho\sigma}{\rm Tr}\left[\bar{B}\sigma^{\mu\nu}\{ u^\rho, D^\sigma B \}\right]\ , \label{LOLVbaryon}%
\end{eqnarray}
\label{LOLV}%
\end{subequations}
where $\tilde{W}^{\mu\nu\rho\sigma} = \epsilon^{\mu\nu\alpha\beta}W_{\alpha\beta}^{\ \ \ \rho\sigma}$. Additional operators with the correct symmetry properties can be constructed at the present chiral order. However, they can all be shown to be redundant up to higher-order terms, using the leading-order equations of motion \cite{Fettes:2000gb,LVchiPT1}, or by using symmetries of the Lorentz indices of the LV coefficients. We omitted any pure pion terms involving $W^{\mu\nu\rho\sigma}$, since they are at least two orders higher in the chiral expansion than the term in Eq.~\eqref{LOLVmeson}. In the final two terms of Eq.~\eqref{LOLVbaryon} we included a factor $1/m_0$ such that NDA designates all LECs $r_1,\ldots,r_4$ to be of order $\mathcal{O}(1)$.

For applications to experimental observations, the explicit lowest order operators in terms of the physical pion and nucleon fields are the most likely to be relevant. They can be found directly from Eqs.~\eqref{LOLS} and \eqref{LOLV} and are given by
\begin{subequations}
\begin{eqnarray}
\mathcal{L}_{\pi} &=& \frac{1}{2}(\partial^\mu\boldsymbol{\pi})\cdot(\partial_\mu\boldsymbol{\pi}) - \frac{1}{2}m_\pi^2\boldsymbol{\pi}^2 \ , \\
\mathcal{L}_{\pi N} &=& \bar{N}\left[i\slashed{D} - m_N - \frac{g_A}{2F_\pi}\gamma^\mu\gamma^5(\boldsymbol{\tau}\cdot \partial_\mu \boldsymbol{\pi})\right]N\ , \\
\mathcal{L}^{k_G}_\pi &=& \frac{r_1}{2}k_{\mu\nu}(\partial^\mu\boldsymbol{\pi})\cdot(\partial^\nu\boldsymbol{\pi})\ , \label{LOLVopion} \\
\mathcal{L}^{k_G}_{\pi N} &=& i r_2 k_{\mu\nu}\bar{N}\gamma^\mu \partial^\nu N + i \frac{r_3 + r_4}{2m_N} \tilde{W}_{\mu\nu\rho\sigma}\bar{N}\sigma^{\mu\nu} (\boldsymbol{\tau}\cdot \partial^\rho \boldsymbol{\pi}) \partial^\sigma N\ , \label{LOLVnucleon}%
\end{eqnarray}
\end{subequations}
where $N = (p,n)^T$ is the nucleon doublet, $\boldsymbol{\tau}\cdot \boldsymbol{\pi} = {\scriptsize \left(\begin{array}{cc} \pi^0 & \sqrt{2}\pi^+ \\ \sqrt{2}\pi^- & -\pi^0 \end{array}\right)}$, and $F_\pi = 92.4\ {\rm MeV}$ is the pion decay constant. Strictly speaking, the LECs for these two-flavor operators are not the same as the ones in Eqs.~\eqref{LOLS} and \eqref{LOLV}, since they receive correction when one integrates out the heavier particles. A determination of the relation between the two sets of LV LECs requires a matching of the two- and three-flavor theories, along the lines of Refs.~\cite{Gas84, Mai}, where relations for the LS LECs can be found. This lies outside the scope of the present considerations. However, since we cannot determine the LECs for the LV operators better than to an order-of-magnitude level anyway, the distinction between two- and three-flavor LECs for LV is not very relevant at present and we will denote them by the same symbol.

An important observation is that the fully traceless tensor $W^{\mu\nu\rho\sigma}$ does not contribute to any kinetic pion or nucleon terms (nor to any other kinetic terms in Eqs.~\eqref{LOlightmeson} and \eqref{LObaryon}). This is reminiscent of a different LVC that was considered in Ref.~\cite{LVchiPT1}. As in that paper, also here it has great consequences for the way limits can be set on $W^{\mu\nu\rho\sigma}$. As we will see, for $k^{\mu\nu}$ we can set limits using the very precise bounds that follow from clock-comparison experiments \cite{datatables}. However, to leading order, these do not pertain to $W^{\mu\nu\rho\sigma}$, because it does not contribute to properties of free protons and neutrons.

\section{Power counting and the heavy-baryon approach}
\label{sec:pc}

A consistent power-counting scheme is necessary to turn the obtained effective theory into a practical tool. We have to know which (loop) diagrams contribute if a certain level of precision is required. To quantify this, one first defines the chiral index $\Delta$, which represents the importance of an operator in the Lagrangian \cite{foundation, Gas84}. In terms of this chiral index a chiral dimension $\nu$ is defined, which specifies the significance of a renormalized Feynman diagram. A diagram of chiral dimension $\nu$ will contribute at order $\mathcal{O}(q^\nu)$, where $q$ is a small quantity of the order of the pion mass. 

In the Lorentz invariant case, the chiral index for operators with at most two baryon fields is given by
\begin{equation}
\Delta = d + f/2 - 2 \ ,
\label{chiralindexLS}
\end{equation}
where $f \leq 2$ counts the number of baryon fields and $d$ is determined by the number of (covariant) derivatives plus twice the number of light-quark masses (since $m_q$ is proportional to $m_\pi^2$). For a LV operator, we use the same definition of the chiral index. This does not directly account for the presence of the small LVC. However, since the coefficients for LV must be heavily suppressed, the LS contributions will essentially always dominate over the LV ones, at least for energies that are relevant in the present context. Therefore, one never needs to compare chiral indices of LV interactions to those of LS interactions.

A generic diagram will now contribute at the following chiral order \cite{Weinbergvol2}:
\begin{equation}
\nu = 2N_L + I_B -N_B + 2 + \sum_i \Delta_i\ ,
\end{equation} 
where $N_L$, $I_B$, and $N_B$ are the number of independent loops, internal baryon lines, and the total number of baryon vertices, respectively, while $i$ runs over the different interactions that contribute to the diagram. For diagrams with exactly one baryon in the initial and final state, it holds that $N_B = I_B +1$ (there are no closed fermion loops in the low-energy EFT) and $\nu$ becomes
\begin{equation}
\nu = 2N_L + 1 + \sum_i \Delta_i\ ,
\label{chiraldim}
\end{equation} 
Again, with a LV insertion, the diagram will be suppressed with respect to any diagram without such an insertion and we can use the same definition of $\nu$ for LV diagrams.

\begin{figure}
\centering
\includegraphics[scale = 0.8]{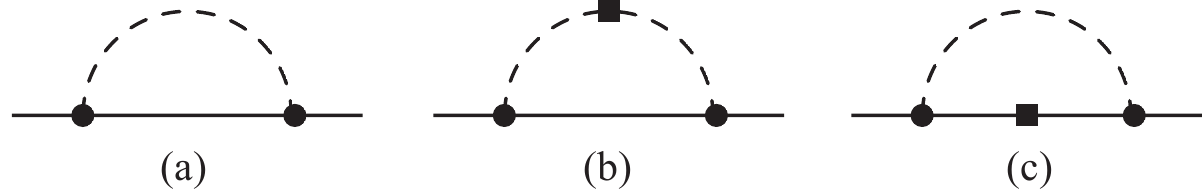}
\caption{Three loop diagrams that contribute to the nucleon self energy. The dots represent conventional $\chi$PT vertices, while the squares represent (different) LV insertions into the pion and nucleon propagators. The solid (dashed) lines are nucleon (pion) propagators.} 
\label{fig:nuclself}
\end{figure}

It is well-known that the diagrams in relativistic meson-baryon theory only obey the power-counting in Eq.~\eqref{chiraldim} if the theory is properly renormalized \cite{Gasser:1987rb}. Otherwise loop calculations will receive contributions of order $m_0$, which is not a small quantity, in fact $m_0/\Lambda_\chi = \mathcal{O}(1)$. These contributions will upset the power counting in Eq.~\eqref{chiraldim}. This holds for LS, as well as LV loop diagrams. For example, the nucleon self energy will receive a contribution from the diagram in Fig.~\ref{fig:nuclself}(b), which involves a LV insertion in the pion propagator, originating from Eq.~\eqref{LOLVopion}. The power counting predicts that this diagram will start to contribute at order $\mathcal{O}(q^\nu) = \mathcal{O}(q^3)$. However, if we calculate the value of the diagram using dimensional regularization, we find that it gives a term
\begin{equation}
\Sigma_{\rm LV} (p^2 = m_{\rm ph}^2) = \left(\frac{\tilde{g}_A^2 m_\pi^3}{16 \pi m_N^2 \tilde{F}_\pi^2}-\frac{\tilde{g}_A^2 m_\pi^2}{16\pi^2 m_N\tilde{F}_\pi^2}\right)r_1 k^{\mu\nu}p_\mu p_\nu + \cdots\ ,
\label{LVnucleoncontr}
\end{equation}
where $p^\mu$ is the nucleon momentum, $m_{\rm ph}$ denotes the physical nucleon mass (to order $q^3$), the tildes indicate renormalized quantities and the dots represent other LV contributions (not necessarily of higher order). We employed the modified minimal substraction scheme of chiral perturbation theory \cite{Gas84}, commonly denoted by $\widetilde{{\rm MS}}$, and for simplicity took the renormalization parameter $\mu = m_N$. The first term in parentheses indeed is of order $\mathcal{O}(q^3)$, as one would expect from Eq.~\eqref{chiraldim}. However, the second term is of order $\mathcal{O}(q^2)$ and thus does not obey the assumed power counting. This already happens in the LS case, where a similar contribution to the nucleon mass appears when the $\widetilde{{\rm MS}}$ scheme is employed in the relativistic theory to calculate the nucleon self energy \cite{Gasser:1987rb}. The power counting can be made consistent by absorbing additional finite terms by counter terms. Infrared regularization \cite{Becher} and the extended on-mass-shell scheme \cite{eoms} are examples of a systematic application of such an approach. However, here we will employ a different approach called heavy-baryon chiral perturbation theory (HB$\chi$PT) \cite{hbchipt}.

The fact that the power-counting is upset in the relativistic theory in the $\widetilde{{\rm MS}}$ scheme, can be traced to the fact that the baryon mass is not small, i.e. time derivatives of the (static) heavy-baryon fields are of order $m_0/\Lambda_\chi = \mathcal{O}(1)$ (while all other terms in the baryon covariant derivative are of order $\mathcal{O}(q)$). To remedy this, in the heavy-baryon formalism the nucleon momentum is usually separated into a large and a small piece like $p^\mu = m_0 v^\mu + k^\mu$, where $v^\mu$ represents a fixed baryon velocity, which obeys $v^2 = 1$ and $k^\mu$ is a small residual momentum. For the LS case, $k^\mu$ also parametrizes how far the nucleon is off-shell, since $p^2 = m_0^2$ if $k = 0$. However, in the LV case, the dispersion relation of the baryons, which follows from Eqs.~\eqref{LObaryon} and~\eqref{LOLVbaryon}, is $\tilde{p}^2 = m_0^2$, with $\tilde{p}^\mu = p^\mu + r_2 k^{\mu\nu}p_\nu$. It is therefore more convenient to define the momentum separation by
\begin{equation}
\tilde{p}^\mu = m_0 v^\mu + \tilde{k}^\mu\ ,
\end{equation}
where $v^2 =1$ still holds, $\tilde{k}^\mu = k^\mu + r_2 k^{\mu\nu}k_\nu$, and $k^\mu$ remains to be a small residual momentum. It follows that $v\cdot \tilde{k} = -\frac{\tilde{k}^2}{2m_0}$. We then define a new heavy-baryon field by
\begin{equation}
B_v = \frac{1}{2}(1 + \slashed{v})e^{i m_0 \hat{v}^\mu x_\mu}B\ ,
\end{equation}
with $\hat{v}^\mu$ defined such that $(\eta^{\mu\nu} + r_2 k^{\mu\nu})\hat{v}_\nu = v^\mu$, i.e. to LO in LV $\hat{v}^\mu = (\eta^{\mu\nu} - r_2 k^{\mu\nu})v_\nu$. Derivatives of these fields will give the small residual momentum such that all derivatives can be counted as $\mathcal{O}(q)$. The baryon propagator no longer contains the large baryon mass. This causes loop diagrams to obey the power counting in Eq.~\eqref{chiraldim} without absorbing any finite terms by counter terms. Additionally, in HB$\chi$PT, the Dirac matrices can be eliminated in favor of the simpler velocity $v^\mu$ and the covariant spin vector $S^\mu = \frac{i}{2}\gamma^5\sigma^{\mu\nu}v_\nu$ with $S = (0,\vec{\Sigma}/2)$, $\vec{\Sigma} = \gamma^5\gamma^0\vec{\gamma}$, for $v = (1,\vec{0})$. The lowest-order Lagrangian for the heavy baryon fields, corresponding to Eqs.~\eqref{LObaryon} and \eqref{LOLVbaryon}, becomes
\begin{subequations}
\begin{eqnarray}
\mathcal{L}_{HB} &=& {\rm Tr}\left[\bar{B}(i v\cdot \partial) B\right] + D\,{\rm Tr}\left[\bar{B}S^\mu \{u_\mu, B\}\right] + F\, {\rm Tr}\left[\bar{B} S^\mu [u_\mu,B]\right] +\cdots\ , \\
\mathcal{L}_{HB}^{k_G} &=& i r_2 k_{\mu\nu} {\rm Tr}\left[\bar{B}v^\mu \partial^\nu B\right] + 2r_3 W_{\mu\nu\rho\sigma}v^\mu v^\sigma {\rm Tr} \left[\bar{B} S^\nu [u^\rho , B]\right] \notag \\  && + 2r_4 W_{\mu\nu\rho\sigma}v^\mu v^\sigma {\rm Tr} \left[\bar{B} S^\nu \{u^\rho , B\}\right] + \cdots\ ,
\end{eqnarray}
\end{subequations}
where we dropped the subscript $v$ on $B_v$. We only kept the leading term for each LVC and the dots represent the higher-order terms and terms with more pions. The propagator of the baryon field becomes $i/(v\cdot \tilde{k})$, which indeed no longer contains a contribution from $m_0$. In terms of nucleon and pion fields, the HB$\chi$PT Lagrangian is
\begin{subequations}
\begin{eqnarray}
\mathcal{L}_{HB} &=& \bar{N}(i v\cdot \partial) N - \frac{g_A}{F_\pi} \bar{N} S^\mu ( \boldsymbol{\tau}\cdot \partial_\mu \boldsymbol{\pi} ) N +\cdots\ , \\
\mathcal{L}_{HB}^{k_G} &=& i r_2 k_{\mu\nu} \bar{N} v^\mu \partial^\nu N + 2(r_3+r_4) W_{\mu\nu\rho\sigma}v^\mu v^\sigma \bar{N} S^\nu(\tau\cdot \partial^\rho \boldsymbol{\pi}) N \notag + \cdots\ ,
\end{eqnarray}
\end{subequations}
where we again only kept the LO terms with the minimum number of pions. 

If we now calculate the contribution of the three diagrams in Fig.~\ref{fig:nuclself}, using the new HB$\chi$PT Lagrangian, we get
\begin{equation}
\Sigma = \frac{3g_A^2 m_\pi^3}{32 \pi F_\pi^2}\left(1 - \left(r_2 - \frac{5}{6}r_1\right)k^{\mu\nu}v_\mu v_\nu\right)\ ,
\end{equation}
which only contains terms of order $\mathcal{O}(q^3)$. This shows that in HB$\chi$PT these diagrams thus obey the power-counting rule in Eq.~\eqref{chiraldim} (at least to the present order), as expected.

\section{Limits from nucleon observables}
\label{sec:nucleonlimits}

The best limits on the semi-traceless part of $k_G^{\mu\nu\rho\sigma}$ come from the fact that it contributes to the effective $c^{\mu\nu}$ parameter for the proton and the neutron. In other words, the first operator in Eq.~\eqref{LOLVnucleon} has the form $ic^{\mu\nu}\bar{\psi}\gamma_\mu\partial_\nu \psi$ and this operator has been studied intensively for the cases that $\psi$ represents the proton or the neutron. The bounds on the components of the neutron and proton $c^{\mu\nu}$ translate almost directly to bounds on the corresponding components of $k^{\mu\nu}$. 

One has to keep in mind, however, that the operator in Eq.~\eqref{LOLVnucleon} contains a LEC $r_2$ whose size can only be estimated by NDA. Moreover, the effective $c^{\mu\nu}$ parameters for the proton and neutron will receive additional contributions from other LV coefficients with the same symmetry properties, in particular from several quark parameters, discussed in Ref.~\cite{LVchiPT2}. It is impossible to completely disentangle the contributions from different coefficients, using just the proton and neutron bounds. The best one can do is obtain a bound on the isospin even (odd) part of $c_{\mu\nu}$ by considering the sum (difference) of the neutron and proton coefficients ($k^{\mu\nu}$ contributes to the isospin even part).

On the other hand, it seems hard to imagine that the contributions from different LVCs conspire to cancel to the level of the stringent proton and neutron bounds. Especially because they all come with LECs that are not related by symmetry arguments. To be conservative, we have therefore set an order of magnitude bound on the components of $k^{\mu\nu}$ that is two orders of magnitude weaker than the best bound on the corresponding proton or neutron parameter. The results are summarized in Table~\ref{tab:kmunubounds}. In fact, only nine of the ten components in the table are independent, since we did not incorporate the tracelessness of $k^{\mu\nu}$.

\begin{table}[t!]
\centering
\begin{tabular}{llc}
\hline\hline
 Tensor component \qquad &  Limit\qquad\qquad & Ref. \\
\hline
$k^{TT}$ & $10^{-21}$ & \cite{Coleman, Kostelecky:2013rta} \\
$k^{TJ}$ & $10^{-19}$ & \cite{Wolf} \\
$k^{JK}$ & $10^{-27}$ & \cite{Smiciklas} \\
$k^{XX}, k^{YY}$ & $10^{-27}$ & \cite{Smiciklas} \\
$k^{ZZ}$ & $10^{-20}$ & \cite{Wolf} \\
\hline
\end{tabular}
\caption{Order-of-magnitude bounds on the LV components of $k^{\mu\nu} = \eta_{\alpha\beta}k_G^{\alpha\mu\beta\nu}$ in the Sun-centered inertial reference frame \cite{datatables}, with $J,K \in \{X,Y,Z\}$. In the right-most column we reference the papers where the corresponding bounds on $c^{\mu\nu}$ for the proton or the neutron were obtained.}
\label{tab:kmunubounds}%
\end{table}

\section{Limits from photon observables}
\label{sec:photonlimits}

In this section, we look at the fully traceless part of $k_G^{\mu\nu\rho\sigma}$, defined in Eq.~\eqref{decomposition}. As can be seen from Eqs.~\eqref{LOLVmeson} and \eqref{LOLVbaryon} there are no kinetic terms in the lowest-order Lagragian when one does not include external fields. However, upon inclusion of electromagnetic fields $W^{\mu\nu\rho\sigma}$ induces the operator
\begin{equation}
\mathcal{L}_{EM} = r_F W_{\mu\nu\rho\sigma}F^{\mu\nu}F^{\rho\sigma}\ ,
\label{LVphoton}%
\end{equation}
with $F^{\mu\nu}$ the photon field strength. The lowest-order Feynman diagram that induces this operator involves a quark-loop in the photon propagator, where a gluon is exchanged between the internal quark lines. Therefore, the NDA estimate for the LEC $r_F$ is given by $r_F = \mathcal{O}(\alpha/(4\pi))$, with $\alpha$ the fine-structure constant. The operator in Eq.~\eqref{LVphoton} has exactly the same form as a photon operator \cite{sme} that involves the parameter $k_F^{\mu\nu\rho\sigma}$. The fully traceless part of this coefficient, defined as in Eq.~\eqref{decomposition}, causes birefringence of light in vacuum \cite{photon1}. By investigating the light from distant gamma-ray bursts (GRBs), very stringent limits have been set on the birefringent part of $k_F^{\mu\nu\rho\sigma}$, which we denote here by $W_F^{\mu\nu\rho\sigma}$. We see now that these bounds are actually bounds on $W_F^{\mu\nu\rho\sigma} + r_F W^{\mu\nu\rho\sigma}$. Because of the symmetry properties in Eq.~\eqref{symmetries} these are the only CPT-even mSME coefficients that, to leading order in LV, contribute to birefringent effects in photons \cite{note1}. 

Using bounds on birefringent photon coefficients \cite{GRBconstraints} one can thus obtain bounds on certain combinations of components of $W^{\mu\nu\rho\sigma}$. We conservatively estimate these bounds to be five orders of magnitude weaker than the limits on $k_F^{\mu\nu\rho\sigma}$, i.e. three orders to account for $r_F = \mathcal{O}(\alpha/(4\pi))$ and two orders for the uncertainty in $r_F$ and partial cancellations between coefficients. The resulting bounds are collected in Table~\ref{tab:Wbounds}.

\begin{table}[t!]
\centering
\begin{tabular}{cc}
\hline\hline
Tensor components \qquad\qquad &  Limit \\
\hline
$\frac{1}{2}W^{TYXZ}, \frac{1}{2}W^{TXYZ},  W^{TXXY}, W^{TXXZ}, W^{TYXY}$ & $10^{-34}$ \\
$W^{TYTY}, W^{TZTZ}, W^{TXTY}, W^{TXTZ}, W^{TYTZ}$ & $10^{-35}$ \\
\hline
\end{tabular}
\caption{Order-of-magnitude bounds on the LV components of $W^{\mu\nu\rho\sigma}$ in the Sun-centered inertial reference frame \cite{datatables}, obtained by comparing to results in Ref.~\cite{GRBconstraints}.}
\label{tab:Wbounds}%
\end{table}
These limits are not independent, in fact, they essentially come from just three measurements. However, any further cancellation between the different components seems unlikely. Using Eq.~\eqref{levicivdif} one can easily translate Table~\ref{tab:Wbounds} to the corresponding limits on $k_G^{\mu\nu\rho\sigma}$. One finds that of the components of $W^{\mu\nu\rho\sigma}$ in Table~\ref{tab:Wbounds}, the top row corresponds to $k_G^a = k_G^1,k_G^2,k_G^8,k_G^9,k_G^{10}$, respectively, while the bottom row corresponds to $k_G^3,\ldots,k_G^7$, respectively, with 
\begin{eqnarray}
k_G^a &=& \big(k_G^{0213},k_G^{0123},k_G^{0202}-k_G^{1313},k_G^{0303}-k_G^{1212},k_G^{0102}+k_G^{1323},k_G^{0103}-k_G^{1223},k_G^{0203}+k_G^{1213}, \notag \\
&& \quad k_G^{0112}+k_G^{0323},k_G^{0113}-k_G^{0223},k_G^{0212}-k_G^{0313}\big)\ ,
\end{eqnarray}
defined analogously to $k^a$ for the photon \cite{photon1}.

\section{Potential improvements}
\label{sec:improvements}

The limits in Tables~\ref{tab:kmunubounds} and \ref{tab:Wbounds} are already quite strict. In fact, they can seem more than sufficient, when one compares them to a reasonable guess for the size of the dimensionless LVCs: $M_{ew}/M_{pl} \simeq 10^{-16}$ with $M_{ew}$ and $M_{pl}$ the electroweak scale and the Planck scale, respectively. However, one does not know what mechanism, if any, would induce these operators and what the associated mass scales are. There are even models where the LV parameters scale with some power of the temperature of the universe \cite{deCesare:2014dga}. Also, all bounds on $k^{\mu\nu}$ are based on one operator and therefore depend on one (renormalized) LEC. Similarly, all bounds on $W^{\mu\nu\rho\sigma}$ depend on $r_F$ (plus loop corrections). It is desirable to get bounds from different effective operators that involve the same LVCs, but different (renormalized) LECs. We consider one option for each coefficient in the following.

\subsection{$k^{\mu\nu}$ and Cherenkov-like pion emission}

In addition to the nucleon operator in Eq.~\eqref{LOLVnucleon}, the semi-traceless part of $k_G^{\mu\nu\rho\sigma}$ also induces the kinetic pion operator in Eq.~\eqref{LOLVopion}. Such a pion operator has been studied before on several occasions \cite{pions}. One of the potential observational consequences of this operator is that it induces an effective refractive index for the vacuum, in the sense that the maximal attainable velocity of the pion will be smaller or larger than the speed of light. If the coefficients have the correct sign, then Cherenkov-like processes can occur, e.g. protons with an energy above some threshold $E_{th}$ will start emitting (neutral) pions until their energy falls below $E_{th}$. An easy way to see this is by realizing that such a process requires the pion to have a spacelike momentum (for simplicity we assume that the proton has a conventional kinetic term). The LV dispersion relation for the pion is then given by $p^2 + r_1 k^{\mu\nu}p_\mu p_\nu - m_\pi^2 = 0$ and therefore the threshold condition becomes
\begin{equation}
|\vec{p}|^2 > \frac{m_\pi^2}{r_1 k^{\mu\nu}\hat{p}_\mu \hat{p}_\nu}\ ,
\end{equation}
with $\hat{p}^\mu = p^\mu/|\vec{p}|$. Clearly, this is only a physical threshold if $r_1 k^{\mu\nu}\hat{p}_\mu \hat{p}_\nu > 0$ and therefore one can never obtain a complete set of limits from Cherenkov-like processes, because the `wrong' sign for the coefficients will not allow for such processes, no matter how large the coefficients are. 

The best sensitivity we can get for the LVC comes from considering ultra-high-energy cosmic-ray protons, which are seen to arrive on Earth with energies above several tens of EeV's from more or less all directions \cite{uhecrdetected}. This means that we can obtain a (one-sided) sensitivity for $k^{\mu\nu}$ of
\begin{equation}
r_1 k^{\mu\nu}\hat{p}_\mu\hat{p}_\mu \lesssim 10^{-23}\ .
\end{equation}
This has the potential of improving at least some of the limits in Tab.~\ref{tab:kmunubounds}.

Notice that actually obtaining limits from Cherenkov-like processes requires some theoretical work. Strictly speaking, one has to calculate the actual emission rate and demonstrate that it does not vanish. But more importantly, one has to verify that the theory with spacelike momenta can be made consistent, since spacelike momenta correspond to negative energies in some observer frames (notably the restframe of the decaying particle). For a different LVC it has been shown that such a theory can nevertheless be quantized and does not contain runaway stability issues \cite{covquantphot}. For the present parameter this remains to be shown, however.

\subsection{$W^{\mu\nu\rho\sigma}$ and the nucleon-nucleon potential}

The fully traceless part of $k_G^{\mu\nu\rho\sigma}$ does not appear in any kinetic term for baryons or light mesons. The lowest-order term allowed by all the symmetries is the one in Eq.~\eqref{LOLVnucleon}, which is a pion-nucleon interaction term. A similar situation was identified in Ref.~\cite{LVchiPT1} for a different LVC. As in Ref.~\cite{LVchiPT1}, we can calculate the nucleon-nucleon potential that follows from considering one-pion exchange between the two nucleons, with one of the vertices originating form the LV $\pi N$ operator in Eq.~\eqref{LOLVnucleon}. The resulting potential is given by
\begin{equation}
V_{\rm LV} = -\frac{8g_A}{F_0}(r_3 + r_4)W^{0i0j}\boldsymbol{\tau}_1 \cdot \boldsymbol{\tau}_2 \frac{(\sigma_1^i \sigma_2^m + \sigma_1^m \sigma_2^i)q^j q^m}{\boldsymbol{q}^2 + m_\pi^2}\ ,
\label{NNpotential}
\end{equation}
where $\boldsymbol{\sigma}_{1,2}$ ($\boldsymbol{\tau}_{1,2}$) are spin (isospin) operators corresponding the interacting nucleons and $\boldsymbol{q} = \boldsymbol{p} - \boldsymbol{p}'$ is the momentum transfer that flows from nucleon 1 to nucleon 2, while $\boldsymbol{p}$ and $\boldsymbol{p}'$ are the relative momenta of the incoming and outgoing nucleon pair in the center-of-mass frame. As usual, the Latin indices run only over spatial directions.

We thus see that $W^{\mu\nu\rho\sigma}$ induces an isospin even two-body operator in the nucleon-nucleon potential. We leave the detailed study of this operator for future work. Obviously, one should expect physical consequences of such a term in nuclear systems with two or more nucleons. For example clock-comparison experiments will most likely be able to provide limits on several components of $W^{\mu\nu\rho\sigma}$. However, a multipole decomposition of $W^{0i0j}$ shows that it only has parts with an angular momentum quantum number of $l=2$. And therefore, at least to leading order, $W^{0i0j}$ will contribute only to clock-comparison experiments involving nuclei with a nuclear spin of $I\geq 1$. The best corresponding sensitivity for LV was obtained in a co-magnetometer experiment involving $^{133}$Cs, which has $I = 7/2$ \cite{Peck}. In that experiment a bound in the order of $10^{-32}\ {\rm GeV}$ on a dimensionful LV parameter was set. Naively we thus expect a bound in the order of $10^{-32}$ on $(r_3+r_4)W^{0i0j}$ from such experiments. 

A related possibility is to study Eq.~\eqref{NNpotential} in the context of the spin precession of the deuteron or other light nuclei in a storage ring \cite{storagering}. Especially the deuteron has the advantage that one does not have to assume a nuclear model to calculate the physical observables. Studying the sidereal variation of the spin-precession frequency of the deuteron could provide limits on $W^{\mu\nu\rho\sigma}$ that are complementary to those quoted in Table~\ref{tab:Wbounds} and to potential bounds from clock-comparison experiments. Most likely, such experiments will not improve on the results in Table~\ref{tab:Wbounds}, however, they have the added benefit that they will be laboratory bounds, which generally involve less assumptions than astrophysical bounds, such as those in Table~\ref{tab:Wbounds}.

\section{Summary and conclusion}
\label{sec:conclusion}

In this paper, we constructed the chiral effective Lagrangian that is induced by the pure gluon LV operator of mass dimension four, which is part of the mSME Lagrangian. We wrote down the dominant operators in the context of a three-flavor $SU(3)$ as well as a two-flavor $SU(2)$ formalism. Relations between the LECs in these two formalisms are yet to be obtained, but are not very relevant for obtaining order-of-magnitude limits on the LVCs. We developed the power-counting rules and showed that in a relativistic meson-baryon theory in the $\widetilde{{\rm MS}}$ scheme, loop contributions to the LVCs upset the power counting, analogous to LS contributions in the conventional theory. To deal with this situation, we wrote down the dominant operators in heavy-baryon chiral perturbation theory and used them to calculate the LV contribution to the nucleon self energy to order $\mathcal{O}(q^3)$.

The symmetries of the mSME gluon operator constrain the form of the chiral effective operators in such a way that kinetic hadron terms can be written down for nine of the nineteen independent components of $k_G^{\mu\nu\rho\sigma}$. We showed that these can be bounded by clock-comparison experiments. The resulting limits are collected in Tab.~\ref{tab:kmunubounds}. We also suggested that additional and improved bounds can be obtained by considering Cherenkov-like pion emission by high-energy protons. 

From our constructed chiral effective Lagrangian we concluded that bounds on the remaining 10 components of $k_G^{\mu\nu\rho\sigma}$, collected in $W^{\mu\nu\rho\sigma}$, cannot be obtained from considerations of free nucleon properties. Therefore, presently available analyses of clock-comparison experiments do not pertain to these components of $k_G^{\mu\nu\rho\sigma}$. On the other hand, $W^{\mu\nu\rho\sigma}$ does induce a photon operator that causes the birefringence of light in vacuum. Therefore we were able to translate existing photon bounds to bounds on $W^{\mu\nu\rho\sigma}$. These are collected in Table~\ref{tab:Wbounds}. Additional and complementary bounds on the fully traceless part of $k_G^{\mu\nu\rho\sigma}$ can most likely be obtained by considering the effect of the nucleon-nucleon potential in Eq.~\eqref{NNpotential} on clock-comparison experiments and storage ring experiments involving the deuteron.

\acknowledgments
We thank J. de Vries and R. Potting for helpful suggestions.
This work is supported by the Funda\c c\~ao para a Ci\^encia e a Tecnologia of Portugal (FCT) through projects UID/FIS/00099/2013 and SFRH/BPD/101403/2014 and program POPH/FSE.


\begin{thebibliography}{99}
\bibitem{note1}
  One could argue that there is also a contribution coming from a $W$-boson loop. This would involve a contribution from $k_W^{\mu\nu\rho\sigma}$ and would be suppressed by a factor $q^4/M_W^4 \simeq 10^{-12}$, where $q \simeq 10^{-4}\ {\rm GeV}$ is the typical energy of photons coming from GRBs. Additionally, there are contributions from a mass-dimension three CPT-odd coefficient. However, for this LVC the birefringence is energy independent and therefore the studies of GRBs are not sensitive to it. This CPT-odd coefficient has been bounded by looking at the isotropy of the CMB polarization and determined to be of order $k^\mu_{AF} < 10^{-43}\ {\rm GeV}$ \cite{datatables}.
\end{thebibliography}
\end{document}